\documentclass[%
 reprint,
superscriptaddress,
 amsmath,amssymb,
 aps,
 prl
] {revtex4-1}

\usepackage{graphicx}% Include figure files
\usepackage{dcolumn}% Align table columns on decimal point
\usepackage{bm}% bold math
\usepackage[]{units}
\usepackage{braket}

\begin{document}

%\preprint{APS/123-QED}

\title{Resilient entangling gates for trapped ions}% Force line breaks with \\
\author{A. E. Webb}
\affiliation{Department of Physics and Astronomy, University of Sussex, Brighton, BN1 9QH, UK}
\author{S. C. Webster}
\affiliation{Department of Physics and Astronomy, University of Sussex, Brighton, BN1 9QH, UK}
\author{S. Collingbourne}
\affiliation{QOLS, Blackett Laboratory, Imperial College London, London, SW7 2BW, UK}
\author{D. Bretaud}
\affiliation{Department of Physics and Astronomy, University of Sussex, Brighton, BN1 9QH, UK}
\affiliation{QOLS, Blackett Laboratory, Imperial College London, London, SW7 2BW, UK}
\author{A. M. Lawrence}
\affiliation{Department of Physics and Astronomy, University of Sussex, Brighton, BN1 9QH, UK}
\affiliation{QOLS, Blackett Laboratory, Imperial College London, London, SW7 2BW, UK}
\author{S. Weidt}
\affiliation{Department of Physics and Astronomy, University of Sussex, Brighton, BN1 9QH, UK}
\author{F. Mintert}
\affiliation{QOLS, Blackett Laboratory, Imperial College London, London, SW7 2BW, UK}
\author{W. K. Hensinger}
\thanks{Author to whom correspondence should be addressed. w.k.hensinger@sussex.ac.uk}
\affiliation{Department of Physics and Astronomy, University of Sussex, Brighton, BN1 9QH, UK}

\date{\today}

\begin{abstract}

Constructing a large scale ion trap quantum processor will require entangling gate operations that are robust in the presence of noise and experimental imperfection. We experimentally demonstrate how a new type of M\o lmer-S\o rensen gate protects against infidelity caused by heating of the motional mode used during the gate. Furthermore, we show how the same technique simultaneously provides significant protection against slow fluctuations and mis-sets in
the secular frequency. Since this parameter sensitivity is worsened in cases where the ions are not ground state cooled, our method provides a path towards relaxing ion cooling requirements in practical realisations of quantum computing and simulation. 

\end{abstract}

\maketitle

Building a quantum processor capable of solving some of the most complex real-world problems will require both a large number of qubits, and the ability to accurately perform gate operations on these qubits. While such gate operations have been demonstrated on pairs of carefully controlled ions with high fidelity \cite{16:ballance, 16:gaebler}, maintaining such fidelities as systems scale towards a large quantum computer will require more robust operations. Many proposed trapped ion quantum processors will require large numbers of ions to be trapped close to the surface of a microfabricated chip \cite{02:kielpinski, 14:monroe, 17:lekitsch, 17:bermudez}, which can cause increased gate infidelities due to heating and dephasing of the ions' motion caused by voltage fluctuations in the electrodes of the chip \cite{16:talukdar} - the heating rate scales unforgivingly with distance as approximately $d^{-4}$ \cite{06:deslauriers}. In addition, it is likely that there will be slowly changing variations in experimental parameters, differing both from position to position on the chip and drifting in time, which will be difficult to fully characterise and correct for during the operation of the processor. This problem is exacerbated when the initial mean excitation of the motional mode, $\bar{n}$, is higher. This could, for example, occur as a result of heating during shuttling processes which form a core part of a number of architectures for a large scale quantum computer \cite{02:kielpinski, 17:lekitsch}. A quantum processor thus requires gate operations that provide low error rates not just under ideal conditions but that are resilient enough to be successfully implemented in realistic experimental environments. 

The two-qubit M{\o}lmer-S{\o}rensen (MS) gate is one of a class of trapped ion gates that operate by state-dependent coherent excitation and de-excitation of a motional mode of a pair of ions during the gate operation \cite{PhysRevLett.82.1971, PhysRevA.62.022311, 03:leibfriedb}, and this motional excitation can be represented as a circular path in phase space. S{\o}rensen and M{\o}lmer discussed how the effect of heating could be reduced by performing multiple smaller circles in phase space \cite{PhysRevA.62.022311}. While this is effective at reducing the impact of heating, the gate time scales as the square root of the number of loops. Hayes et al.\ experimentally demonstrated a similar technique as a method of reducing the effect of a `symmetric' detuning error, such as could be caused by an incorrectly measured trap frequency \cite{12:hayes}. 

Noise suppression can be effectively achieved with less impact on the gate time by tracing out more complicated paths in phase space \cite{16:Haddadfarshi, 15:green, 18:leung}. Recent theoretical work proposed a method whereby the infidelity due to heating can be significantly reduced with smaller impact on gate time than by just performing multiple smaller loops \cite{16:Haddadfarshi}. Here, we experimentally demonstrate this effect, using a pair of trapped $^{171}$Yb$^+$ ions. We then build upon this result to show that these same paths also dramatically increase the resilience of the gate to errors caused by symmetric detuning errors. We demonstrate how this resilience becomes particularly significant in the case of the mode used for the gate not being cooled close to the ground state.

By off-resonantly driving the red and blue motional sidebands of a pair of ions coupled to a common motional mode the MS Hamiltonian,
\begin{equation}
H_{\rm MS}=\frac{\hbar \delta}{4}\hat{S}_{x}(\hat{a}^\dagger e^{i\delta t}+\hat{a} e^{-i \delta t}),
\label{eq:Hms}
\end{equation}
can be realised, where $\hat{S}_{x}=\hat{\sigma}_{x1} + \hat{\sigma}_{x2}$ is the sum of the $\hat{\sigma}_{x}$ matrices for the two ions, $\hat{a}^\dagger$ and $\hat{a}$ are the motional mode raising and lower operators respectively, and $\delta$ is the magnitude of the detuning from the red and blue sidebands. When the driving fields are applied for a time $\tau=2\pi/\delta$, the pair of ions undergo the unitary transformation
 \begin{equation}
 U_{\rm MS}=\exp{\left[i\frac{\pi }{4}\hat{\sigma}_{x1}\hat{\sigma}_{x2}\right]}
 \end{equation}
 which can produce a maximally entangled state from an initial product state. 

In practice, there are a number of mechanisms by which the fidelity of the MS gate will be reduced from one. Here we consider two sources of infidelity: a dephasing of the gate caused by heating of the motional mode during the gate process, and an incorrect phase pickup and residual entanglement of the qubits with the motional mode caused by a symmetric detuning frequency error. This symmetric detuning occurs as a result of the magnitude of the gate detuning $\delta$ being set incorrectly by the same amount for both ions, $\Delta$, for instance due to a drift in trap frequency.  

If the total infidelity is small, the fidelity of the gate can be expressed as a sum of independent infidelities ${\rm F}=1-({\rm E}_{\rm h}+{\rm E}_\Delta+{\rm E}_{\rm oth})$, where $\rm E_h$ is the infidelity due to heating, $\rm E_\Delta$ the infidelity due to symmetric detuning error, and $\rm E_{oth}$ is the sum of any other infidelities, which we henceforth do not consider. Since the infidelities are small we will consider the infidelities only to the leading order in either heating rate or detuning error. The errors due to non-zero heating rate and symmetric detuning error are then
 \begin{equation} \label{eq:MSsym}
 {\rm E}_{\rm h}=\frac{\pi\dot{\bar{n}}}{\delta}, \ \ {\rm E}_{\Delta}=\left(\frac{3}{4}+\bar{n}\right)\pi^2\left(\frac{\Delta}{\delta}\right)^2
 \end{equation}
respectively, where $\dot{\bar{n}}$ is the heating rate and $\Delta$ is the error in gate detuning (see Supplemental Material \cite{SM}\nocite{18:randall, 13:noek, 09:burrell}). Note that the error due to incorrect detuning is also dependent on the mean excitation of the motional mode at the start of the gate $\bar{n}$ (we assume the phonon distribution to be thermal) - if the ion is `hotter' (larger $\bar{n}$) at the start of the gate operation, it becomes much more sensitive to parameter errors.

\begin{figure}
\centering
\includegraphics[width=73mm]{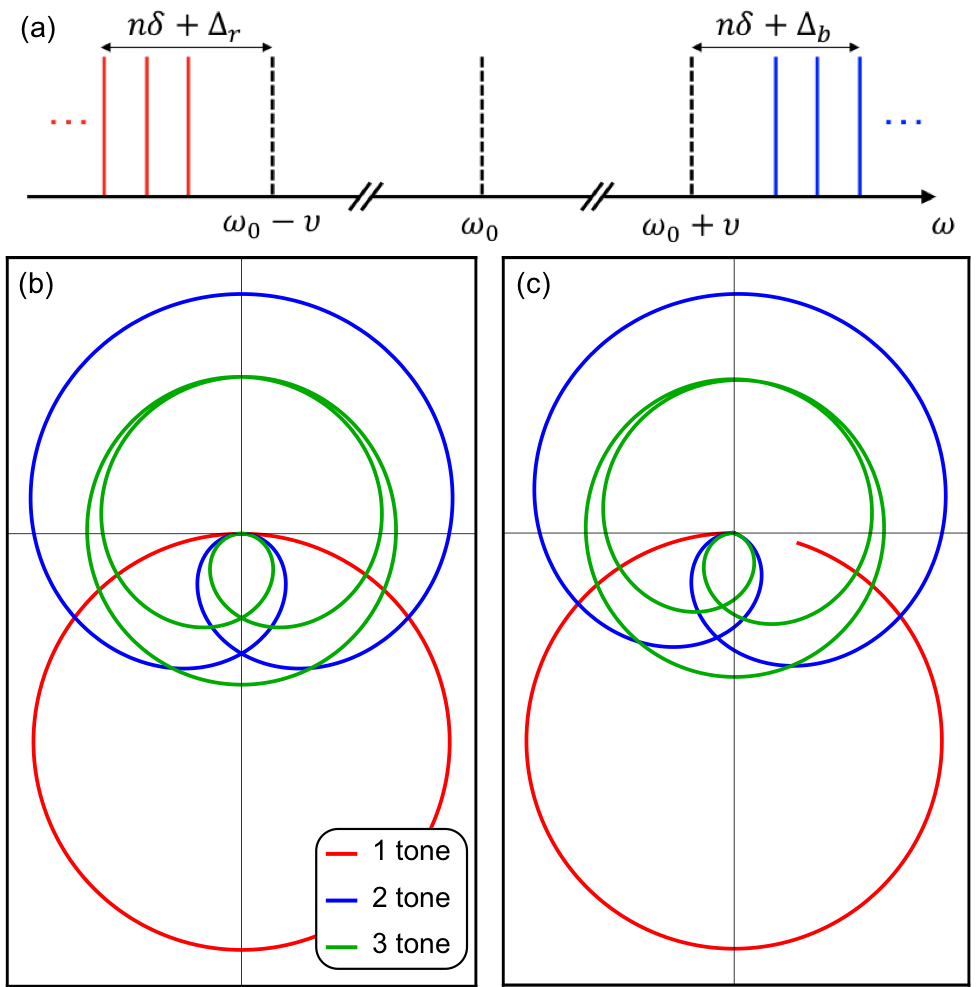}
\caption{(a) Energy level diagram showing multi-tone gate fields detuned from the correct gate detuning $\delta$ by $\Delta_r$ and $\Delta_b$ for the red and blue sidebands respectively. MTMS gates provide protection against errors of the form $\Delta_r=\Delta_b=\Delta$. (b) Phase space trajectories for one (red), two (blue), and three (green) tone gates. Unlike the single tone case, the average displacement $\langle\alpha(t)\rangle=0$ for two or more tones. In addition, as the number of tones increases, $\langle|\alpha(t)|^2\rangle$ becomes smaller. The effect of this reduction in squared displacement is to reduce adverse effect of heating on gate fidelity. (c) For a single tone gate, a symmetric detuning error ($\delta/\Delta=0.05$) results in both incomplete loops in phase space, causing error due to the residual entanglement between the spin and motional states of the qubit, and incorrect phase accumulation.  For the two and three tone gates, the loops come much closer to completion (visually indistinguishable from closed loops), effectively eliminating residual qubit-motion entanglement as a contribution to the infidelity, and although not visually obvious, the phase picked up also becomes closer to the ideal. Both of these effects lead to a reduction in sensitivity to symmetric detuning errors.}
\label{fig:phase}
\end{figure}

Haddadfarshi et al.\ introduced a method to reduce the effect of heating and dephasing of the motional mode on the gate fidelity \cite{16:Haddadfarshi}. They considered a multi-tone generalisation of the MS gate (MTMS), where instead of driving each ion sideband with a single field, MTMS gates use $N$ fields or tones to drive each sideband at detunings $\delta_j=j \delta$ with $\{j=1,...N\}$ as shown in figure \ref{fig:phase}(a), and each tone's strength is given by coefficients $c_j$. The Hamiltonian thus becomes 
\begin{equation}
H_{\rm MS}=\hbar \delta\hat{S}_{x}\sum_{j=1}^N c_j(\hat{a}^\dagger e^{ij\delta t} +\hat{a} e^{-ij\delta t}).
\end{equation}
The condition to produce a maximally entangling unitary constrains the values of the coefficients $c_j$ to be $\sum_{j=1}^N\frac{c_j^2}{j}=\frac{1}{16}$, which corresponds to a standard single tone MS gate having a coefficient $c_1=\frac{1}{4}$.

The effect of any MS type Hamiltonian is to excite the motion of different spin components, causing them to selectively acquire a phase, before (ideally) returning them to their initial motional state. This excitation can be considered as a time varying displacement in a rotating phase-space, and the effect of heating depends quadratically on the magnitude of this displacement during the gate. Haddadfarshi et al.\ found that the best reduction in the effect of heating of the mode over the course of the gate is found when the average phase space displacement is zero, and average squared phase space displacement is minimised over the course of the gate. This corresponds to parameters where $\sum_{k=1}^N\frac{c_k}{k}=0$, and $\sum_{k=1}^{N}\frac{|c_k|^2}{k^2}$ is minimised (see Supplemental Material \cite{SM} for more detail). The effect of using the MTMS gate on the phase-space trajectories can be seen in figure \ref{fig:phase}(b). 

For MTMS gates, the heating rate defining the infidelity is modified by a factor given by (see Supplemental Material \cite{SM})
\begin{equation}
\dot{\bar{n}}_{\rm MT} = 8\left( \sum_{k=1}^N\frac{c_k^2}{k^2} + \left(\sum_{k=1}^N\frac{c_k}{k}\right)^2\right)\dot{\bar{n}}.
\end{equation}
The minimisation conditions mean that this can be understood as a smaller effective heating rate. For $N=\{1,2,3\}$, then $\dot{\bar{n}}_{\rm MT}=\{1, 1/3, 1/5.19\}\times\dot{\bar{n}}$ respectively.

We show here that MTMS gates also protect against errors due to an incorrect symmetric detuning. In this case, the MTMS Hamiltonian is modified to become
\begin{equation}
H_{\rm MS}=\hbar \delta\hat{S}_{x}\sum_{j=1}^N c_j(\hat{a}^\dagger e^{i(j\delta+\Delta)t} +\hat{a} e^{-i(j\delta+\Delta) t})
\end{equation}
By expanding the fidelity of the gate in powers of the fractional symmetric detuning error, $\Delta/\delta$, we obtain the same set of constraints on the coefficients, $c_j$, as was obtained when minimising the effect of motional decoherence, so the gate is protected against both sources of infidelity (see Supplemental Material \cite{SM}).

The infidelity due to symmetric detuning error of the optimised MTMS gates to leading order in $\Delta/\delta$ is given by
\begin{align}
{\rm E_{\Delta}^{MT}} &\simeq 16\pi^2\left(\frac{\Delta}{\delta}\right)^2\left(\sum_{j=1}^N\frac{c_j^2}{j^2}\right)^2 \\
& = \frac{1}{36}\pi^2\left(\frac{\Delta}{\delta}\right)^2 \simeq 0.028\pi^2\left(\frac{\Delta}{\delta}\right)^2 & (N=2) \\
 & = \frac{39-12\sqrt{3}}{1936}\pi^2\left(\frac{\Delta}{\delta}\right)^2 \simeq 0.0094\pi^2\left(\frac{\Delta}{\delta}\right)^2 & (N=3) 
\end{align}
The sensitivity to $\Delta$ is significantly reduced for two and three tone gates compared with the standard MS gate (eq.\ \ref{eq:MSsym}). The infidelity is also independent of the initial distribution of motional states, unlike eq.\ \ref{eq:MSsym}. This is because the effect of residual qubit-motional entanglement on the fidelity is zero to this order of $\Delta/\delta$, the infidelity being completely due to the incorrect phase being acquired during the gate. This may be of particular interest when gates are combined with shuttling operations which induce heating \cite{02:kielpinski,17:lekitsch}. Figure 1(c) shows gates with symmetric detuning error $\delta/\Delta=0.05$. The single tone gate produces an obviously incomplete loop. For the MTMS gates, while the detuning produces a rotation of the phase space paths, the two paths appear indistinguishable from closed loops - the two tone gate comes almost 70 times closer to completion than the single tone, the three tone gate almost 360 times closer.\\

%\subsubsection*{experimental}

We demonstrate this technique experimentally using a pair of $^{171}$Yb$^+$ ions \cite{PhysRevA.83.013406}. The hyperfine ground state is driven using microwave and radiofrequency radiation, and a magnetic field gradient of \unit[23.6(3)]{T/m} is generated using permanent magnets to enable the requisite coupling between the internal spin and collective motional modes \cite{01:mintert, 15:lake}. The ions are decoupled from magnetic field noise using a dressed state system \cite{Timoney:2011aa, 13:webster, 15:randall}. A pair of microwave fields for each ion couple $\ket{^2S_{1/2},F=0}\equiv \ket{0}$ with $\ket{^2S_{1/2},F=1,m_F=-1} \equiv \ket{-1}$  and $\ket{^2S_{1/2},F=1,m_F=-1} \equiv \ket{+1}$  and, in the interaction picture, this gives three well protected states including $(\ket{+1}-\ket{-1})/\sqrt{2} \equiv \ket{D}$. The pair of states $\ket{D}$ and $\ket{^2S_{1/2},F=1,m_F=0} \equiv \Ket{0'}$ gives a well protected qubit with transition frequencies \unit[11.0]{MHz} and \unit[13.9]{MHz} for each ion respectively and a coherence time of \unit[500]{ms}.

A maximally entangled Bell state is created and analysed for standard single tone ($N=1$) and two tone ($N=2$) MS gates, since moving from one to two tones should show the largest improvement in gate robustness. The single tone MS gate procedure is detailed in more detail in \cite{16:weidt}. The gate was performed on the stretch mode of the ions, with frequency $\nu/2\pi= \unit[461]{kHz} $ which gives an effective Lamb-Dicke parameter of $\eta=0.004$. Single and two tone gates with the same gate time $\tau$ and detuning $\delta$ were compared. The single tone gate uses a pair of gate fields per ion, each of carrier Rabi frequency $\Omega_0/2\pi=\unit[36]{kHz}$, and, since $\delta=2\eta\Omega_0$ the detuning is $\delta/2\pi=\unit[292]{Hz}$ and the gate time is $\tau=2 \pi / \delta=\unit[3.42]{ms}$. The two tone gate uses two pairs of gate fields per ion, with the Rabi frequencies of the two tones in each pair being $\Omega_1=-0.576 \Omega_0$ at $\delta$ and $\Omega_2=1.152 \Omega_0$ at $2\delta$, corresponding to $c_1=-0.144$ and $c_2=0.288$. The beating between the two tones produces a time-varying Rabi frequency, and thus a time varying Stark shift that we compensate for by changing the gate field detunings during the gate operation - see Supplemental Material \cite{SM} for more information.  Before performing the gate operation the stretch mode was sideband cooled to an initial temperature of $\bar{n}\approx0.1$ \cite{PhysRevLett.115.013002}. 

The fidelity of Bell-state production is found by measuring selected elements of the density matrix, specifically the total population in the states $\ket{0'0'}$ and $\ket{DD}$,  and the coherence between these two states \cite{16:weidt}. A maximum likelihood method was used to determine these two values, as well as the errors in their measurement (see Supplemental Material \cite{SM}).

\begin{figure}
\centering
\includegraphics[width=80mm]{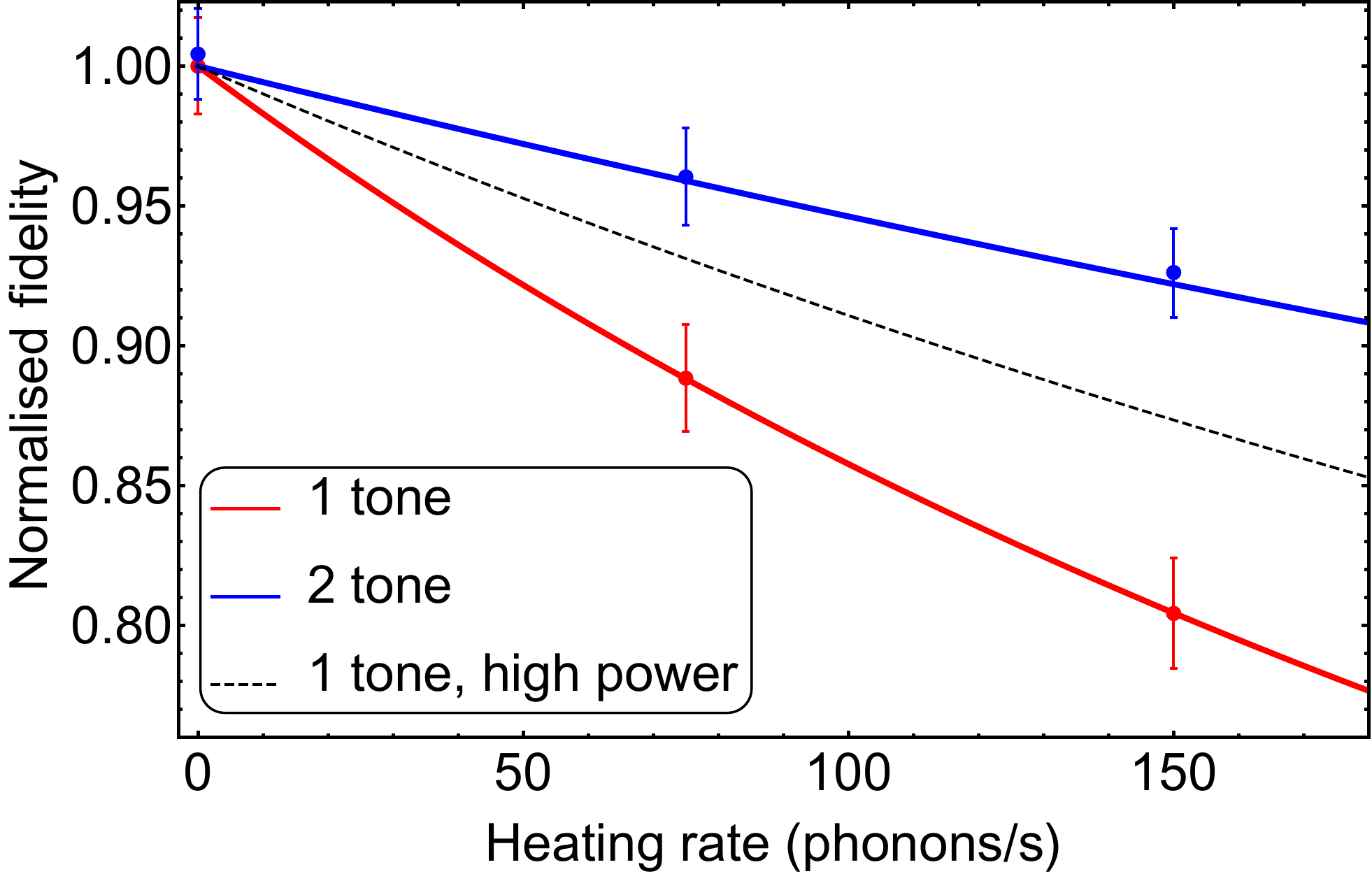}
\caption{Infidelities due to heating are reduced by moving from a single to a two tone MS gate, shown in red and blue respectively. Both gates are of duration $\tau=\unit[3.4]{ms}$. Solid lines are the results of a numerical simulation, and three experimental points are shown for each type of gate where the heating rate has been increased artificially through noise injection. Fidelities are normalised to the fidelity at the single tone fidelity measured at the lowest heating rate (0.94(2)) to account for other errors \cite{figurenote}. The two tone gate requires a higher peak Rabi frequency for a given gate duration. Numerical simulation of a faster single tone gate with this higher Rabi frequency is shown in a dashed black line, showing the two-tone gate is still superior.
}
\label{fig:heating}
\end{figure}

To demonstrate the effectiveness of the MTMS technique for protection against heating, the heating rate was artificially increased. Randomly generated noise with a flat amplitude spectrum over a bandwidth of \unit[20]{kHz} centred around the secular frequency, $\nu$, was capacitively coupled onto an endcap DC trap electrode, and the heating rate controlled by changing the amplitude of this noise. Heating rates with no added noise, and for two different amplitudes of artificial noise, were measured by introducing a varying time delay after sideband cooling and measuring the temperature of the ion using sideband spectroscopy. Figure \ref{fig:heating} shows the gate fidelity as a function of these three heating rates for both single and two tone gates.The solid lines are the results of a numerical simulation of the master equation with appropriate Lindblad operators to model heating, the results of which show good agreement with the theoretical values for fidelity given by equation S.7 of the Supplemental Material \cite{SM}. The dashed line is the result of a numerical simulation for a faster single tone gate at a higher power, as defined by the peak Rabi frequency used for the two tone gate, and demonstrates that two tone gates still exhibit lower error due to heating. No increase in fidelity is observed for the two tone gate at no induced heating compared to the single tone gate due to the small contribution to overall infidelity from heating, smaller than the measurement uncertainty. The measured infidelity at no induced heating is thought to be largely a result of dephasing and depolarising, and parameter mis-set primarily of the form $\Delta_b\neq\Delta_r$, where $\Delta_r$ and $\Delta_b$ are the detuning errors on each sideband (see Figure 1(a)). Methods to protect against this error exist \cite{17:Manovitz} and combining these with MTMS techniques may offer a solution.

\begin{figure}
\centering
\includegraphics[width=80mm]{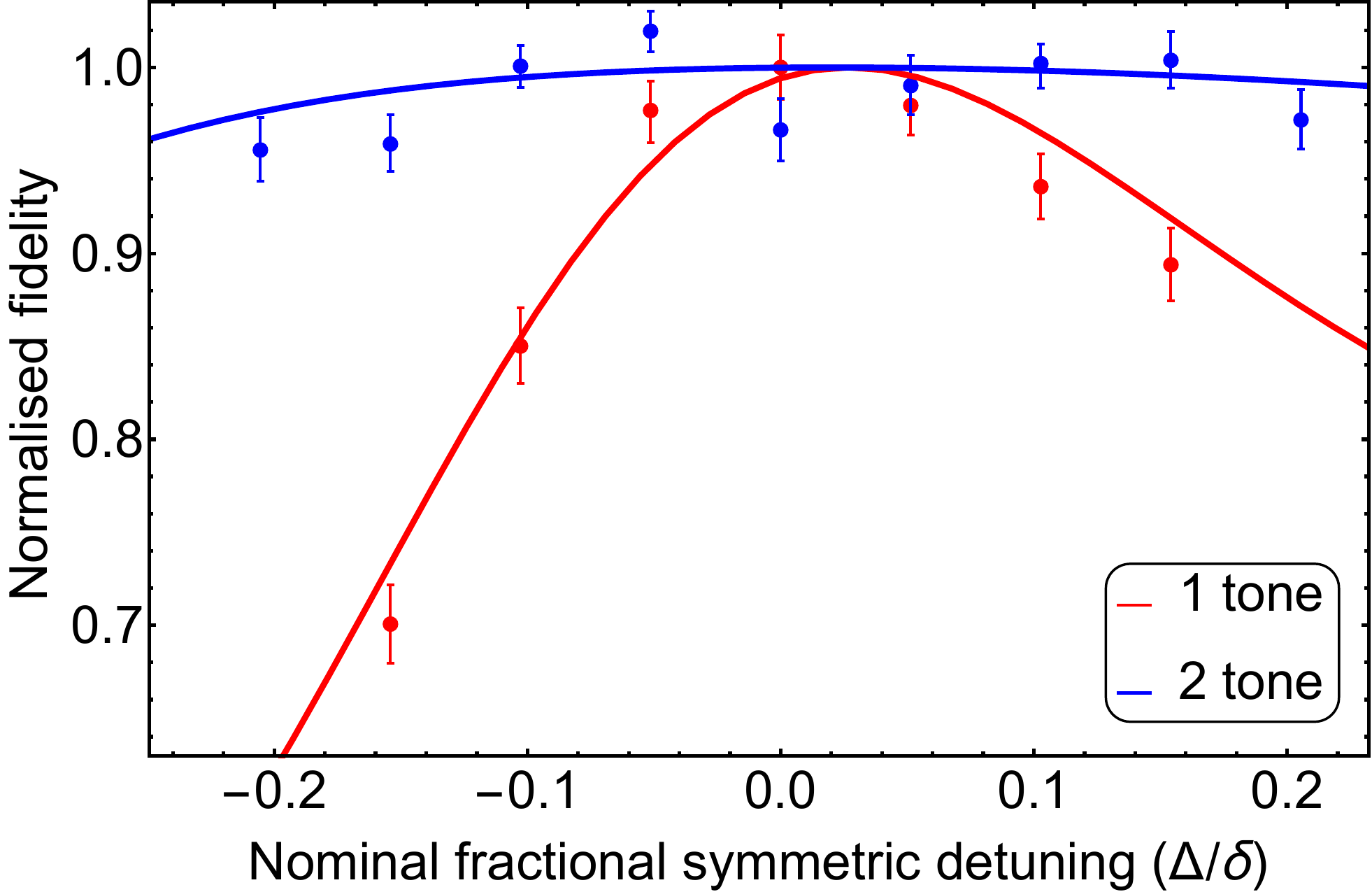}
\caption{The effect of symmetric detuning error is significantly reduced by moving to two tones, shown here for a gate duration of $\tau=\unit[3.4]{ms}$. The experimental fidelities for non-zero nominal detuning errors are calculated with respect to the Bell state defined by zero nominal detuning, to account for any phase shift induced by the error. Solid lines are from numerical simulations, but with an offset applied to detuning error to account for any error in determining the trap frequency. Experimentally determined fidelities are normalised to the single tone fidelity with zero nominal detuning error (0.96(2)) to account for other infidelities. The symmetric detuning offset  was fitted using the single tone theory curve to account for the uncertainty in the initial setting of experimental parameters.
}
\label{fig:sym}
\end{figure}

In order to demonstrate robustness to symmetric detuning errors, a symmetric detuning error of up to $\pm0.2\delta$ was added to the nominal zero error detuning \footnote{Due to a.c.\ Stark shift caused by the gate fields causing a symmetric detuning offset, and the difficulty of accurately measuring the carrier Rabi frequency $\Omega_0$ directly, we have to experimentally determine the correct value of the detuning $\delta$ for a given gate time $\tau$}. Results are shown in figure \ref{fig:sym}, where again solid lines show the result of numerical simulation. A clear consistency between simulation and experimental results can be seen, demonstrating strong protection against both heating and detuning errors obtained by using a two tone rather than the standard single tone MS gate. 

Since the symmetric detuning error for multi-tone gates no longer exhibits any dependence on the initial mean excitation of the motional mode of the ions to first order in $\Delta/\delta$, this also opens up the possibility of performing gates at higher $\bar{n}$ which can for example be reached by Doppler cooling.  A multi-tone gate of fidelity 0.851(9) has been demonstrated at an initial thermal state with $\bar{n}=53(4)$, compared to a single tone fidelity of 0.50(5), as shown in figure 4. The dominant infidelity of the MTMS gate is expected to be due to detuning errors of the form $\Delta_r \neq \Delta_b$, which remain sensitive to $\bar{n}$. 

\begin{figure}
\centering
\includegraphics[width=80mm]{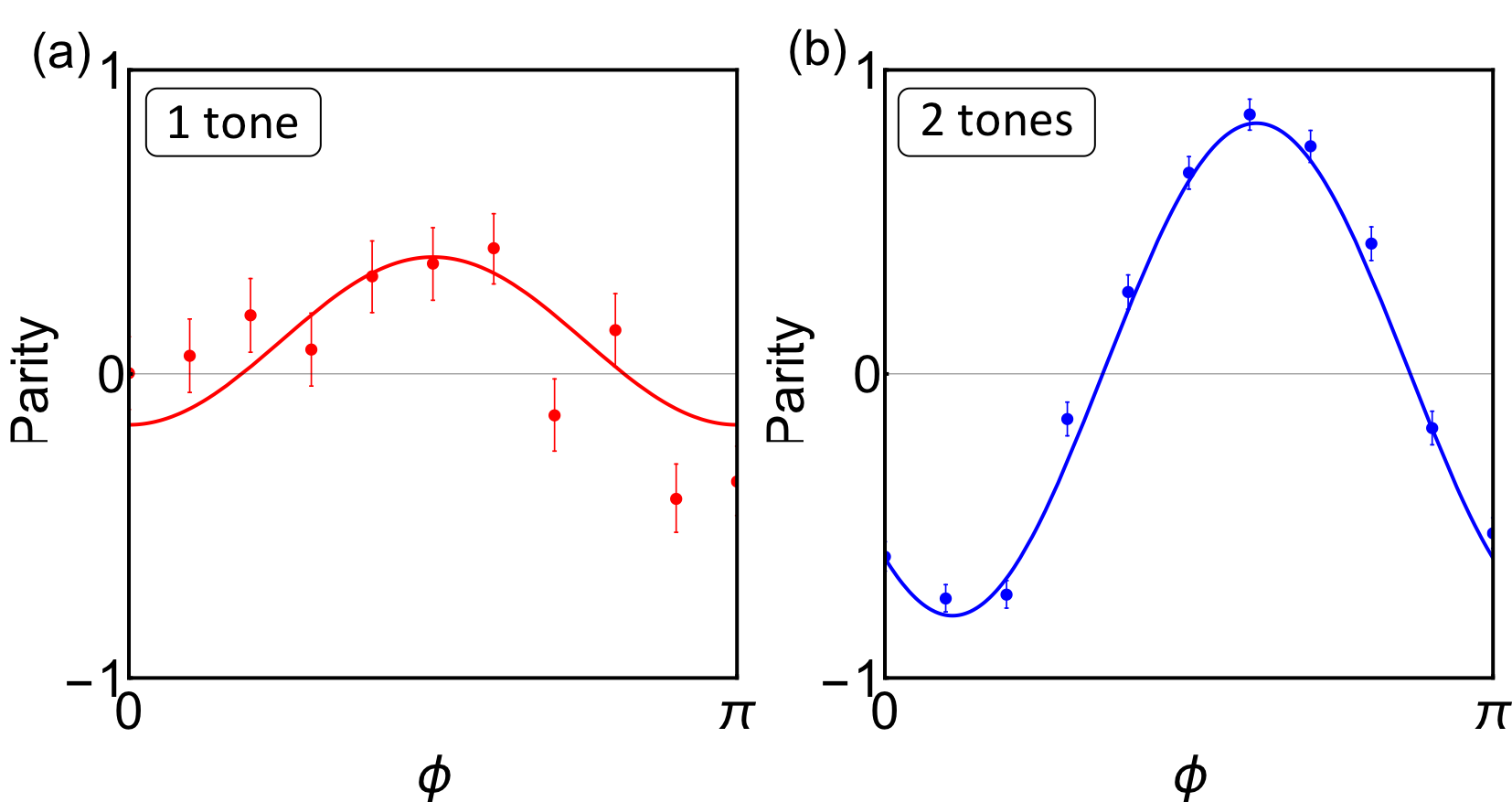}
\caption{Parity curves at an initial motional mode of $\bar{n}=53(4)$ for a single (red) and two tone (blue) gate after ions have been  only Doppler cooled. A significant improvement in contrast is seen for the two tone gate, since the dependence of gate infidelity on $\bar{n}$ to first order in $\Delta$ is eliminated by the use of multiple tones.}
\label{fig:hotgates}
\end{figure}

We have shown that for a given gate time, the use of a two tone MS gate instead of a standard single tone MS gate substantially reduces the effect of motional heating, as well as significantly lowering the sensitivity to symmetric detuning errors. This comes at a cost in terms of resources - the peak power required to drive the gate is three times higher, while the average power is $\frac{5}{3}$ times higher. However, provided heating is a significant source of error, MTMS gates should prove a powerful tool, particularly for large scale systems required for quantum computing with comparatively low ion heights and potentially noisier and less stable environments \cite{16:johnson, 14:guise}. 

In addition, we note that MTMS gates also provide protection against fluctuations in the trap frequency caused by Kerr coupling of the stretch mode to the radial mode \cite{08:Roos, 09:Nie}, which is only Doppler cooled, thus alleviating one of the main restrictions on using the stretch mode for two qubit gates. Finally, use of MTMS gates also act to reduce off-resonant excitation of the carrier caused by the gate fields compared to a single tone gate, despite the higher peak power. This is due to the lower initial Rabi frequency and the sinusoidal variation in Rabi frequency acting as a natural pulse shaping to reduce this excitation, opening up the potential for performing faster gates.

After the preparation of the manuscript, we have become aware of related work where similar methods were used to make laser based entangling gates more robust \cite{18:shapira}.
  
\begin{acknowledgments}

The authors thank Farhang Haddadfarshi for helpful discussions. This work is supported by the U.K. Engineering and Physical Sciences Research Council [EP/G007276/1; the U.K. Quantum Technology hub for Networked Quantum Information Technologies (EP/M013243/1) and the U.K. Quantum Technology hub for Sensors and Metrology (EP/M013294/1)], the European Commissions Seventh Framework Programme (FP7/2007-2013) under grant agreement no. 270843 Integrated Quantum Information Technology (iQIT), the Army Research Laboratory under cooperative agreement no. W911NF-12-2-0072, the U.S. Army Research Office under contract no. W911NF-14-2-0106, the European Research Council (Project: Optimal dynamical control of quantum entanglement), and the University of Sussex. This work was supported through a studentship in the Quantum Systems Engineering Skills and Training Hub at Imperial College London funded by the EPSRC (EP/P510257/1).
\end{acknowledgments}

\bibliography{bibpoly}

\begin{thebibliography}{10}
\bibitem{SM}
{S}ee {S}upplemental {M}aterial for further details, which includes {R}efs.
  [17-19].
\bibitem{figurenote}
The gate infidelity is primarily due to dephasing and depolarisation of the
  qubits, and parameter mis-sets of the form \mbox{$\Delta_r\neq\Delta_b$}
  caused by difficult to characterise {S}tark shifts. {W}e did not fully
  compensate errors due to {S}tark shifts, which are the largest contribution
  to the infidelity, in order to accelerate acquisition of data since this work
  is focussed on the mitigation of symmetric detuning and heating induced
  errors, rather than realising a gate of high fidelity.
\end{thebibliography}
\bibliographystyle{unsrt}

\newif\ifarxiv

\arxivtrue

\setcounter{equation}{0}
\renewcommand{\theequation}{S.\arabic{equation}}

\newpage

\renewcommand{\thepage}{Supplemental Material -- \arabic{page}}
\setcounter{page}{1}

\onecolumngrid

\section{\large{SUPPLEMENTAL MATERIAL}}

\twocolumngrid

\subsection{Protection against motional decoherence}

Here we briefly summarise the work presented in \ifarxiv \cite{16:Haddadfarshi} \else [13] \fi to protect the fidelity of an MS gate from the effects of decoherence of the motional mode due to heating or dephasing of the motional mode. The MS Hamiltonian (equation \ifarxiv \ref{eq:Hms}\else 1\fi) can be generalised
\begin{equation}
H_{\rm MS}=\hbar \delta\hat{S}_{x}(a^\dagger f^*(t)+a f(t)),
\label{eq:Hmsgen}
\end{equation}
which by using the Magnus expansion gives a unitary transformation
\begin{equation}
U_{\rm MS}(t)=\exp(-i(F(t)a+F^*(t)a^\dagger)S_x - G(t)S_x^2))
\label{eq:Umsgen}
\end{equation}
where $F(t)=\delta\int_0^t dt'f(t')$ and $G(t)=-\frac{i\delta}{2}\int_0^t dt'(f^*(t')F(t')-f(t')F^*(t'))$. The function $F(t)$ gives, modulo a phase factor, the displacement of the motional state of selected spin states during the gate operation and should be equal to zero at the gate time $\tau$. $G(t)$ gives the amount of phase accumulated during the gate, and should be equal to $\pi/8$ at the gate time. 

The effect of phonon decoherence  is to produce a dephasing which depends on time averages of functions of $f(t)$ over the course of the gate. Haddadfarshi et al.\ found that the effect of heating was most reduced by setting $\langle F(t)\rangle$ to zero, and minimising $\langle|F(t)|^2\rangle$.

To find optimised functions $f(t)$ the function is parameterised as a Fourier series of $N$ different tones
\begin{equation}
f(t) = \sum^N_{j=1}c_j\exp{(ij\delta t)}.
\end{equation}
The condition for a maximally entangling gate gives the constraint $\sum_{k=1}^N\frac{c_k^2}{k}=\frac{1}{16}$, and $\langle F(t)\rangle=0$ means $\sum_{k=1}^N\frac{c_k}{k}=0$. For a two tone $N=2$ gate, up to a phase factor this fixes the values of the coefficients to $c_1=-0.144$ and $c_2=0.288$. For $N\geq2$, the minimisation of $\langle|F(t)|^2\rangle$ corresponds to minimising $\sum_{k=1}^N\frac{|c_k|^2}{k^2}$.

The solution to this minimisation can be shown to be
\begin{equation}
c_j=\frac{jb}{1-j\lambda}
\end{equation}
where
\begin{equation}
b = -\frac{1}{4}\left(\sum_{j=1}^N\frac{j}{(1-j\lambda)^2}\right)^{-\frac{1}{2}}
\end{equation}
and $\lambda$ is the smallest root of the equation
\begin{equation}
\sum_j^m(1-j\lambda)^{-1} = 0
\end{equation}

Values of $c_j$ for $N\leq8$ can be found in \ifarxiv \cite{16:Haddadfarshi}. \else [13]. \fi

\subsection{Infidelity due to heating}

We study the effect of applying multiple tones to the case where the motional decoherence is purely due to heating of the mode, and here we quantify the effect on fidelity of different numbers of tones for a given heating rate. We define the fidelity of the gate ${\rm F}=\left<\Psi_{\rm max}|\rho_{\rm int}|\Psi_{\rm max}\right>$, where $\left |\Psi_{\rm max}\right>$ is the appropriate maximally entangled state, and $\rho_{\rm int}$ is the reduced density matrix of the internal states of the ions.

The infidelity caused by heating can be modelled using a master equation with appropriate relaxation operators \ifarxiv \cite{PhysRevA.62.022311}. \else [10]. \fi In the case where the phonon decoherence is solely due to heating of the mode at a rate $\dot{\bar{n}}$ it can be shown that the gate fidelity is given by the standard result for a single loop MS gate
\begin{equation}
{\rm F}=\frac{1}{8}[3+4\exp{(-\dot{\bar{n}}_{\rm MT}\tau/2)}+\exp{(-\dot{2\bar{n}}_{\rm MT}\tau)}] 
\label{eq:heatingFid}
\end{equation}
but with the heating rate modified by a factor to give an effective heating rate $\dot{\bar{n}}_{\rm MT}$ which depends on the tone coefficients
\begin{equation}
\dot{\bar{n}}_{\rm MT} = 8\left( \sum_{k=1}^N\frac{c_k^2}{k^2} + \left(\sum_{k=1}^N\frac{c_k}{k}\right)^2\right)\dot{\bar{n}}.
\end{equation}
Note that the second term is proportional to $\langle F(t)\rangle$ and is thus zero for $N\geq2$ and the first term is proportional to $\langle|F(t)|^2\rangle$, which is the quantity minimised for $N\geq3$, as would be expected. For $N=\{1,2,3\}$, then $\dot{\bar{n}}_{\rm MT}=\{1, 1/3, 1/5.19\}\times\dot{\bar{n}}$ respectively.

When the infidelity is small, equation \ref{eq:heatingFid} can be simplified by expanding to first order in $\dot{\bar{n}}$:
\begin{equation}
{\rm F} \simeq 1 - \frac{\pi\dot{\bar{n}}_{\rm MT}}{\delta}.
\end{equation}

\subsection{Protection against detuning error}

Here we show how taking the same approach of parameterising the MS Hamiltonian by using multiple harmonic tones can be applied to the problem of detuning error.

In the presence of an arbitrary detuning error $\Delta$, and using the Fourier parameterisation, the MS Hamiltonian (\ref{eq:Hmsgen}) becomes
\begin{equation}
H_{\rm MS}=\hbar \delta\hat{S}_{x}(a^\dagger e^{-i\Delta t} f^*(t)+a\,e^{i\Delta t}f(t))
\label{eq:Hmsdet}
\end{equation}
which by use of the Magnus expansion leads to the following expressions for the functions $F(t)$ and $G(t)$ at the conclusion of the gate:
\begin{eqnarray}
F(\tau)&=&\sum_{j=1}^N\frac{c_j\delta}{i(k\delta+\Delta)}(e^{i\frac{2\pi\Delta}{\delta}}-1)\\
G(\tau)&=&\sum_{j,k=1}^N\frac{c_jc_k\delta^2}{i(k\delta+\Delta)}\left(\frac{2\pi}{\delta}\delta_{jk}-\frac{\sin(\frac{2\pi\Delta}{\delta})}{j\delta+\Delta}\right)
\end{eqnarray}

This results in two different decohering effects when $\Delta\neq0$. The fact that $F(\tau)\neq0$ means there is residual qubit-photon entanglement at the end of the gate and the fact that $G(\tau)\neq\frac{\pi}{8}$ means that the wrong amount of phase is acquired during the gate operation. The fidelity of the gate operation can be expressed in terms of the values of $F(\tau)$ and $G(\tau)$ as
\begin{align}
{\rm F}  =  \frac{3}{8} &+\frac{1}{2}\cos{(4G_\Delta(\tau))}\exp{(-4(\bar{n}+\frac{1}{2})|F(\tau)|^2)} \nonumber \\
& + \frac{1}{8}\exp{(-16(\bar{n}+\frac{1}{2})|F(\tau)|^2)} \label{eq:fullFid}
\end{align}
where $G_\Delta(\tau)=G(\tau)-\frac{\pi}{8}$ is the deviation of $G(\tau)$ from the ideal case.

By expanding $F(\tau)$ in $\Delta$, constraints can be found which eliminate the qubit-phonon coupling to a given order in $\Delta^m$:
\begin{equation}
\left\{ \sum_{k=1}^N\frac{c_k}{k^j}=0 \right\}_{j=1}^m
\end{equation}

Eliminating qubit-phonon coupling to first order in $\Delta$ results in the same constraint as when $\langle F(t)\rangle$ is set to zero, and thus for the two tone case this results in the same coefficients as Haddadfarshi et al.\ found as optimum for protection against heating.

We can see from equation \ref{eq:fullFid} that eliminating qubit-phonon coupling to first order in $\Delta$ removes its effect on fidelity to second order in $\Delta$, which also removes the dependence on fidelity of the initial temperature $\bar{n}$. The fidelity can then be approximated, to leading order in $\Delta$, to be
\begin{eqnarray}
{\rm F} & \simeq & 1 - 4G_\Delta^2(\tau) \nonumber \\
 & \simeq & 1 - 16\pi^2\left(\frac{\Delta}{\delta}\right)^2\left(\sum_{j=1}^N\frac{c_j^2}{j^2}\right)^2.
 \end{eqnarray}

This then means that for $N\geq3$ that minimising the infidelity to leading order in $\Delta$ leads to the same set of constraints as were obtained in minimising the infidelity due to motional decoherence and so for a given value of $N$ the same set of coefficients $c_j$ minimises both forms of infidelity under consideration.

We can then compare the fidelity of the standard single tone MS gate, to the same order in $\Delta$, to the optimised two and three tone MS gates ${\rm F}=1-{\rm E}_\Delta^{\rm MT}$, where:
\begin{align}
{\rm E}_\Delta^{\rm MT} &\simeq  \left(\frac{3}{4}+\bar{n}\right)\pi^2\left(\frac{\Delta}{\delta}\right)^2 & (N=1) \\
&=\frac{1}{36}\pi^2\left(\frac{\Delta}{\delta}\right)^2\simeq 0.028\pi^2\left(\frac{\Delta}{\delta}\right)^2 & (N=2) \\
 &=\frac{39-12\sqrt{3}}{1936}\pi^2\left(\frac{\Delta}{\delta}\right)^2\simeq 0.0094\pi^2\left(\frac{\Delta}{\delta}\right)^2 & (N=3) 
 \end{align}

\subsection{M{\o}lmer-S{\o}rensen gate using a magnetic field gradient}

In the interaction picture, applying radiation fields detuned from the red and blue stretch mode sidebands of a pair of ionic qubits trapped within a magnetic field gradient produces the following Hamiltonian:
\begin{equation}
H_{MS}=\frac{i\hbar \eta \Omega_0}{2}\hat{S}_{y}(a^\dagger e^{i\delta_0 t}-a e^{-i \delta_0 t})
\label{eq:Hmsfull}
\end{equation}
where $\hat{S}_{y}=\hat{\sigma}_{y1}-\hat{\sigma}_{y2}$,  $\delta$ is the detuning, $\eta$ is the Lamb-Dicke parameter, $\Omega_0$ the carrier Rabi frequency, and we have set the phases of the driving field to zero. This differs from the canonical form of the MS Hamiltonian in a number of respects. The first is that $\hat{S}_{y}$ is the difference between the single ion spin operators, caused by the fact we are driving the stretch mode, and the second is the factor of $i$ and sign difference between the $a$ and $a^\dagger$ terms, which is due to the fact that the atom-photon coupling is produced by a combination of long-wavelength radiation and static field gradient, rather than photon momentum from laser light.
 
By applying the radiation fields for a time $\tau=2\pi/\delta$ to a pair of ions whose internal states and motional states are separable, over the course of the gate the motion returns to its initial state, while the internal states are transformed according to
 \begin{equation}
 U=\exp{\left[i\frac{\pi \eta^2 \Omega_0^2}{\delta^2}\hat{\sigma}_{y1}\hat{\sigma}_{y2}\right]}.
 \end{equation}
Note that as we use the stretch mode we perform a phase gate in the $\sigma_y$ basis, rather than the $\sigma_x$ basis described in the main text. The effect of this is to change the phase of the final Bell state produced during the gate. By setting the detuning $\delta=2\eta\Omega_0$ the required entangling two-qubit unitary $U=\exp{\left[i\frac{\pi }{4}\hat{\sigma}_{y1}\hat{\sigma}_{y2}\right]}$ is then obtained. Finally, we can rewrite equation \ref{eq:Hmsfull} in terms of $\delta$:
\begin{equation}
H_{MS}=\frac{i\hbar \delta}{4}\hat{S}_{y}(a^\dagger e^{i\delta t}-a e^{-i \delta t}).
\label{eq:Hms1}
\end{equation}

\subsection{Stark shifts}

Due to the asymmetry of the dressed state system, the gate fields produce an a.c. Stark shift. For single tone gates this is corrected for by shifting the frequency of the gate fields. For two tone gates the beating due to the relatively small frequency separation $\delta$ between the two tones causes an effective sinusoidal amplitude modulation of the gate fields. Since the a.c. Stark shift depends on this amplitude, it is also sinusoidally time-varying, and is compensated for by varying the detuning of the gate fields over the gate time.

\subsection{Fidelity measurement}

We measure the state of the ions by carrying out a fluorescence measurement and detecting the scattered light on a PMT. We then threshold the photon counts from the PMT to obtain three quantities $x_0$, $x_1$ and $x_2$, which are, respectively, the number of times we observe that both ions are in the dark state; exactly one ion is in the bright state; and both ions are in the bright state, where $x_0 + x_1 + x_2 = n$. From these we can estimate the true probabilities for each of these outcomes, by way of a maximum likelihood method \ifarxiv \cite{18:randall}. \else [17]. \fi 

The probabilities to measure each of the three outcomes $p'_0$, $p'_1$ and $p'_2$ differ from the true probabilities $p_0$, $p_1$ and $p_2$ due to errors in the preparation and state detection of the ions. When calibrating the state detection measurements, we typically measure a combined state preparation/ detection fidelity of around $87\%$ (note that near unity detection fidelity could be obtained using appropriate imaging optics and detectors \ifarxiv \cite{13:noek, 09:burrell}). \else [18, 19]). \fi From these calibrations, we can extract a linear map $p'_i = \sum_jP(i|j)p_j$, which relates the probabilities of the measurement outcomes to the true probabilities. (Note that we only need to use $p_1$ and $p_2$, as $p_0$ is constrained by $p_0 + p_1 + p_2 = 1$.) Since the measured counts $x_i$ are distributed according to a multinomial probability distribution, we can find most likely values for the true probabilities $p_i$ by maximising the following log-likelihood function,  
\begin{widetext}
\begin{equation}
f(p_1, p_2) = \log\left(\frac{(n+1)(n+2)n! p'_1(p_1,p_2)^{x_1} p'_2(p_1,p_2)^{x_2} (1-p'_1(p_1,p_2)-p'_2(p_1,p_2))^{n-x_1-x_2}}{x_1!x_2!(n - x_1 - x_2)!}\right),
\end{equation}
\end{widetext}
over the variables $p_1$ and $p_2$. 

In order to measure the gate fidelity we use the parity oscillation method described in \ifarxiv \cite{16:weidt}. \else [26].\fi Here, the parity $\Pi = p_0 + p_2 - p_1$ undergoes oscillations of the form $\Pi = A\cos(2\phi + \phi_0)$ where $\phi$ is the phase of the analysis pulse, and $\phi_0$ the phase of the entangled state. We make $N$ parity measurements $\Pi^i$ at varying analysis pulse phases $\phi^i$, from which we must extract a best fit value of $A$ in order to obtain the Bell state fidelity. This fit is achieved using a maximum likelihood method similar to that described above. This time our measured values are the counts the `even' and `odd' parity states $x_\mathrm{even} = x_0 + x_2$ and $x_\mathrm{odd} = x_1$ where $x_\mathrm{even} + x_\mathrm{odd} = n$. We use the log-likelihood function
\begin{equation}
f = \sum_{i = 1}^N \log\left(\frac{(n+1)n! p'_\mathrm{odd}(p^i_\mathrm{odd})^{x^i_\mathrm{odd}} (1-p'_\mathrm{odd}(p^i_\mathrm{odd}))^{n-x^i_\mathrm{odd}}}{x^i_\mathrm{odd}!(n-x^i_\mathrm{odd})!}\right),
\end{equation}  
where $p^i_\mathrm{odd} = (1 - \Pi^i)/2$ and we are summing over all $N$ data points $x^i_\mathrm{odd}$ in the fit. We can then maximise $f$ with respect to the fit parameters $A$ and $\phi_0$, and use these to calculate the fidelity.

In order to account for any change in the phase of the Bell state produced, the amplitude of the parity fit was multiplied by a a factor $\cos{\Delta\phi}$, where $\Delta \phi=\phi_m-\phi_0$, $\phi_m$ is the phase of the Bell state as measured and $\phi_0$ is the phase of the Bell state expected (measured at either no induced heating or no induced symmetric detuning error).

\end{document}